\documentclass[authoryear,preprint,review,12pt]{elsarticle}

\usepackage{graphicx}
\usepackage{txfonts}
\usepackage[figuresright]{rotating}
\usepackage{bm}
\usepackage{natbib}

\usepackage{hyperref}
\hypersetup{colorlinks=true,
            citecolor=blue,
            linkcolor=blue}

\usepackage[footnotesize]{caption}

\def\stc {Lempo }
\def\vv#1{\bm{#1}}
\def\be {\begin{equation}}
\def\ee {\end{equation}}
\def\vR {\vv{R}}
\def\ddvR{\vv{\ddot R}}
\def\vf {\vv{f}}
\def\vg {\vv{g}}
\def\vh {\vv{h}}
\def\vu {\vv{u}}
\def\vq {\vv{q}}
\def\dvq {\vv{\dot q}}
\def\vr {\vv{r}}
\def\vrkl {\vv{r}_{kl}}
\def\ur {\vv{\hat r}}
\def\urkl {\vv{\hat r}_{kl}}
\def\rkl {r_{kl}}

\def\vF{\vv{F}}
\def\vT {\vv{T}}
\def\dvr{\vv{\dot r}}
\def\ddvr{\vv{\ddot r}}
\def\w{\omega}
\def\vw {\vv{\w}}
\def\tvw {\vv{\tilde \w}}
\def\vL {\vv{L}}
\def\dvL{\vv{\dot L}}

\def\ia{i}
\def\jb{j}
\def\kc{k}
\def\vi {\vv{\ia}}
\def\vj {\vv{\jb}}
\def\vk {\vv{\kc}}
\def\xii {\hat x}
\def\xj {\hat y}
\def\xk {\hat z}

\def\TI {{\bf \cal I}}
\def\TJ {{\bf \cal J}}
\def\SR {{\bf \cal S}}

\def\M {M}
\def\m {m}

\def\tt {\{t\}}

\def\crm{\cr\noalign{\medskip}}

\def\llabel#1{\label{#1}}
\def\nul#1{}

\def\bfx#1{{#1}}

\def\figpath{}

\begin{document}

\begin{frontmatter}

\title{Chaotic dynamics in the (47171) \stc triple system}

\author[lab1,lab2]{Alexandre C. M. Correia}

\address[lab1]{CIDMA, Departamento de F\'isica, Universidade de Aveiro, Campus de Santiago, 3810-193 Aveiro, Portugal}

\address[lab2]{ASD, IMCCE-CNRS UMR8028, Observatoire de Paris, 77 Av. Denfert-Rochereau, 75014 Paris, France}

\begin{abstract}
We investigate the dynamics of the (47171) \stc triple system, also known by 1999\,TC$_{36}$.
We derive a full 3D~$N$-body model that takes into account the orbital and spin evolution of all bodies, which are assumed triaxial ellipsoids.
We show that, for reasonable values of the shapes and rotational periods, the present best fitted orbital solution for the \stc system is chaotic and unstable in short time-scales.
The formation mechanism of this system is unknown, but the orbits can be stabilised when tidal dissipation is taken into account.
The dynamics of the \stc system is very rich, but depends on many parameters that are presently unknown.
A better understanding of this systems thus requires more observations, which also need to be fitted with a complete model like the one presented here.
\end{abstract}

\begin{keyword}
celestial mechanics \sep planetary systems \sep planets and satellites: individual: (47171) \stc
\end{keyword}

\end{frontmatter}

\section{Introduction}

A non-negligible fraction of the small bodies in the solar system are in multiple systems, mostly composed by binaries \citep[e.g.][]{Noll_etal_2008}.
The shapes of these small objects are usually irregular \citep{Lacerda_Jewitt_2007}, resulting in important asymmetries in the gravitational potential.
The dynamics of these objects is thus very rich, as these asymmetries lead to strong spin-orbit coupling, where the rotation rate can be captured in a half-integer commensurability with the mean motion \citep{Colombo_1965, Goldreich_Peale_1966}.
For very eccentric orbits or large axial asymmetries, the rotational libration width of the individual resonances may overlap, and the dynamics becomes chaotic \citep{Wisdom_etal_1984,Wisdom_1987}.
When a third body is added to the problem, the mutual gravitational perturbations also introduce additional spin-orbit resonances at the perturbing frequency \citep{Goldreich_Peale_1967, Correia_etal_2015, Delisle_etal_2017}.

\bfx{ 
The spin and orbital dynamics of small-body binaries has been object of many previous studies.
However, due to the complexity of the spin-orbit interactions, in general these works either focus on the spin or in the orbital dynamics, i.e., they study the spin of a triaxial body around a distant companion \citep[e.g.][]{Batygin_Morbidelli_2015, Naidu_Margot_2015, Jafari_Assadian_2016}, or the motion of a test particle around a triaxial body \citep[e.g.][]{Mysen_Aksnes_2007, Scheeres_2012, Lages_etal_2017}.
Moreover, for simplicity, most studies consider that the spin axis is always normal to the orbital plane, and when a third body is considered, the orbits are also made coplanar. 
}

(47171) \stc (also known by 1999\,TC$_{36}$) is a triple system. 
It is classified as a {\it plutino}, since it is in a 3/2 mean-motion resonance with Neptune, like Pluto.
The primary was discovered in 1999 at the Kitt Peak Observatory \citep{Rubenstein_Strolger_1999}.
A similar size secondary was identified in 2001 from images obtained by the Hubble Space Telescope \citep{Trujillo_Brown_2002}.
Subsequent observations lead to the determination of the orbit of the secondary with a period of  about~50 days  \citep{Margot_etal_2005}.
A third component, also of similar size, was finally discovered in 2007 also using observations from the Hubble Space Telescope \citep{Jacobson_Margot_2007}.
The third body is actually much closer to the primary than the secondary, with an orbital period of only 1.9~days \citep{Benecchi_etal_2010}.
The \stc system can thus be characterised as an inner close binary with an outer circumbinary companion, with all three components being of identical sizes, \bfx{which is unique}. 

The name {\it Lempo} actually refers to the larger component of the inner binary, while the smaller component is named {\it Hiisi}, and the outer circumbinary component is named {\it Paha}.
The best fitted orbits for the \stc system are eccentric ($\sim0.1$ for the inner orbit and $\sim0.3$ for the outer one) and present a mutual inclination of about 10 degrees \citep{Benecchi_etal_2010}.
The three bodies have diameter sizes within $100-300$~km \citep{Mommert_etal_2012}, which is consistent with a large triaxiality.
Therefore, since the two inner components are very close to each other, we expect to observe a strong spin-orbit coupling in this system.

In this paper we derive a full 3D model (for the orbits and spins) that is suitable to describe the motion of a $N-$body system, where all bodies are assumed triaxial ellipsoids (section~\ref{model}).
\bfx{ This model is able to simultaneously handle spin and orbital dynamics without any kind of restrictions}.
We then apply our model to the \stc system in section~\ref{condyn}, and show that the present best fitted solution corresponds to a chaotic system for reasonable values of the unknown triaxiality.
In section~\ref{tidevol} we analyse the impact of tidal evolution on the final evolution of the system.
Finally, in last section we discuss our results.

\section{Model}

\llabel{model}

In this section we derive a very general model that is suited to study a system of $N$-bodies with ellipsoidal shapes. 
Our model is valid in 3D for the orbits and individual spins. 
\bfx{We make no particular assumption on the spin axes}.
We use cartesian inertial coordinates, and quaternions to deal with the rotations.

\llabel{secmodel}

\subsection{Potential of an ellipsoidal body}

We consider an ellipsoidal body of mass $\m$, and chose as reference the cartesian inertial frame ($\vi,\vj,\vk$).
In this frame, the rotational angular velocity and angular momentum vectors of the body are given by $\vw = (\w_\ia, \w_\jb, \w_\kc)$ and $\vL = (L_\ia, L_\jb, L_\kc)$, respectively, which are related through the inertia tensor $\TI$ as
\be
\vL = \TI \cdot \vw  \quad \Leftrightarrow \quad \vw = \TI^{-1} \cdot \vL 
\ ,
 \llabel{151019b}
\ee
where
\be
\TI = 
\left[\begin{array}{rrr} 
I_{11}&  I_{12}& I_{13} \crm
I_{12}&  I_{22}& I_{23} \crm
I_{13}&  I_{23}& I_{33} 
\end{array}\right] \ .
\label{121026c}
\ee

The gravitational potential of the ellipsoidal body at a generic position $\vr$ from its center-of-mass is given by \citep[e.g.,][]{Goldstein_1950}
\be 
V (\vr) = - \frac{G \m}{r} + \frac{3 G}{2 r^3} \left[ \ur \cdot \TI \cdot \ur - \frac{1}{3} \mathrm{tr}(\TI) \right] \ , \llabel{121026b}
\ee
where $G$ is the gravitational constant, $\ur = \vr / r = (\xii, \xj, \xk)$ is the unit vector, and $\mathrm{tr}(\TI) = I_{11} + I_{22} + I_{33} $.
We neglect terms in $(R/r)^3$, where $R$ is the mean radius of the body (quadrupolar approximation).
Adopting the Lagrange polynomial $P_2(x)=(3x^2-1)/2$, we can rewrite the previous potential as
\begin{eqnarray}
V (\vr) = - \frac{G \m}{r} 
+ \frac{G}{r^3}  \Big[ \big(I_{22}-I_{11}\big) P_2 (\xj) + \big(I_{33}-I_{11}\big) P_2 (\xk) && \nonumber \\ 
+ \, 3 \big( I_{12} \xii \xj + I_{13} \xii \xk + I_{23} \xj \xk \big) \!\!\!\!\!\! &  \Big] & \!\!\!\!\!\!
\ . \llabel{151028a}
\end{eqnarray}

\subsection{Point-mass problem}

\llabel{pmp}

We now consider that the ellipsoidal body orbits a point-mass $\M$ located at $\vr$.
The force between the two bodies is easily obtained from the potential energy of the system $ U (\vr) = \M V (\vr) $  (Eq.\,\ref{151028a}) as 
\be
\vF = - \nabla U (\vr) =  \vf(\M, \m, \vr) + \vg(\M, \TI, \vr) + \vh(\M, \TI, \vr) 
\ , \llabel{170911d}
\ee
with
\be
\vf(\M, \m, \vr) = - \frac{G \M \m}{r^3} \vr \ ,  \llabel{170911a}
\ee
\begin{eqnarray}
\vg(\M, \TI, \vr) \!\!\!\!\! & = & \!\!\!\!\!  
\frac{15 G \M}{r^5}  \Big[ \frac{I_{22}-I_{11}}{2} \big(\xj^2 - \frac15 \big)  
+ \frac{I_{33}-I_{11}}{2} \big(\xk^2 - \frac15 \big) \nonumber \\ 
&& \!\!\!\!\! + I_{12} \xii \xj + I_{13} \xii \xk + I_{23} \xj \xk \Big] \vr 
\ , \llabel{170911b}
\end{eqnarray}
\begin{eqnarray}
\vh(\M, \TI, \vr) \!\!\!\!\! & = & \!\!\!\!\! 
 -  \frac{3 G \M}{r^4}  \Big[ \big(I_{22}-I_{11}\big) \xj \vj + \big(I_{33}-I_{11}\big) \xk \vk \nonumber \\ 
 && \!\!\!\!\! + I_{12} (\xii \vj + \xj \vi)  + I_{13} (\xii \vk + \xk \vi) + I_{23} (\xj \vk + \xk \vj) \Big]
\ . \llabel{170911c}
\end{eqnarray}
We thus obtain for the orbital evolution of the system
\be
\ddvr =  \vv{F} / \beta \ ,  \llabel{151028c}
\ee
where $\beta = \M \m/ (\M + \m)$ is the reduced mass.
The spin evolution of the ellipsoidal body can also be obtained from the force, by computing the gravitational torque. In the inertial frame we have: 
\be
\dvL = \vT (\M, \TI, \vr) = - \vr \times \vv{F} = - \vr \times \vh 
\ , \llabel{150626a}
\ee
that is,
\begin{eqnarray}
\vT (\M, \TI, \vr) = \frac{3 G \M}{r^3} \ur \times \Big[ \big(I_{22}-I_{11}\big) \xj \vj + \big(I_{33}-I_{11}\big) \xk \vk && \nonumber \\ 
+ I_{12} (\xii \vj + \xj \vi) + I_{13} (\xii \vk + \xk \vi) + I_{23} (\xj \vk + \xk \vj) \!\!\!\!\!\! &  \Big] & \!\!\!\!\!\!
\ , \llabel{151028d}
\end{eqnarray}
or
\be
\vT =  \frac{3 G \M}{r^3}
\left[\begin{array}{c}  
\big(I_{33}-I_{22}\big) \xj \xk - I_{12} \xii \xk + I_{13} \xii \xj  + I_{23} (\xj^2 - \xk^2)  \crm 
\big(I_{11}-I_{33}\big) \xii \xk + I_{12} \xj \xk + I_{13} (\xk^2 - \xii^2) - I_{23} \xii \xj  \crm 
\big(I_{22}-I_{11}\big) \xii \xj + I_{12} (\xii^2 - \xj^2) - I_{13} \xj \xk + I_{23} \xii \xk
\end{array}\right] 
\ . \llabel{151028e}
\ee

Apart from a sphere, in the inertial frame ($\vi,\vj,\vk$) the inertia tensor (\ref{121026c}) is not constant.
We let $\SR$ be the rotation matrix that allow us to convert any vector $\vu_B$ in a frame attached to the body into the cartesian inertial frame $\vu_I$, such that $\vu_I = \SR \, \vu_B$.
Thus, we have
\be
\TI = \SR \, \TI_B \SR^T
\ , \quad \mathrm{and} \quad
\TI^{-1} = \SR \, \TI_B^{-1} \SR^T   \ ,
\llabel{160127a}
\ee
where $\TI_B$ is the inertia tensor in the body frame.
For principal axis of inertia $\TI_B = \mathrm{diag}(A, B, C)$ and $\TI_B^{-1} = \mathrm{diag}(A^{-1}, B^{-1}, C^{-1})$.
The evolution of $\SR$ over time is given by 
\be
\dot \SR = \tvw \, \SR  \ , \quad \mathrm{and} \quad \dot \SR^T = - \SR^T \tvw  
\ , \llabel{160127b}
\ee
with 
\be
\tvw = \left[\begin{array}{ccc} 
0   &  - \w_\kc   & \w_\jb \crm
\w_\kc   & 0   & - \w_\ia \crm
- \w_\jb   & \w_\ia   & 0
\end{array}\right] 
\ . \llabel{160111b}
\ee
In order to simplify the evolution of $\SR$, a set of generalized coordinates to specify the orientation of the two frames can be used.
Euler angles are a common choice, but they introduce some singularities.
Therefore, here we use quaternions \citep[eg.][]{Kosenko_1998}.
We denote $\vq = (q_0, q_1, q_2, q_3)$ the quaternion that represents the rotation from the body frame to the inertial frame.
Then
\be
\SR = \left[\begin{array}{ccc} 
q_0^2+q_1^2-q_2^2-q_3^2   &  2 (q_1 q_2 - q_0 q_3)   & 2 (q_1 q_3 + q_0 q_2) \crm
2 (q_1 q_2 + q_0 q_3)   & q_0^2-q_1^2+q_2^2-q_3^2   & 2 (q_2 q_3 - q_0 q_1) \crm
2 (q_1 q_3 - q_0 q_2)   & 2 (q_2 q_3 + q_0 q_1)   & q_0^2-q_1^2-q_2^2+q_3^2
\end{array}\right] 
\ , \llabel{160127d}
\ee
and
\be
\dvq = \frac{1}{2} (0,\vw) \cdot \vq 
= \frac{1}{2} \left[\begin{array}{c} 
-\w_\ia q_1 - \w_\jb q_2 - \w_\kc q_3 \crm 
\w_\ia q_0 + \w_\jb q_3 - \w_\kc q_2 \crm 
- \w_\ia q_3 + \w_\jb q_0 + \w_\kc q_1 \crm 
\w_\ia q_2 - \w_\jb q_1 + \w_\kc q_0  
\end{array}\right] 
\ . \llabel{160127c}
\ee

To solve the spin-orbit motion, we need to integrate equations (\ref{151028c}), (\ref{150626a}) and  (\ref{160127c}), using the relations (\ref{151019b}), (\ref{160127a}) and (\ref{160127d}).

\subsection{Two-body problem}

Consider now that two ellipsoidal bodies with masses $\m_0$ and $\m_1$, and inertia tensors $\TI_0$ and $\TI_1$, respectively, orbit around each other at a distance $\vr$ from their centers-of-mass.
The total potential energy can be written from expression (\ref{121026b}) as
\be 
U = - \frac{G \m_0 \m_1}{r} + \frac{3 G}{2 r^3} \left[ \ur \cdot \TJ \cdot \ur - \frac{1}{3} \mathrm{tr}(\TJ) \right] \ , \llabel{170911e}
\ee
with $\TJ = \m_0 \TI_1 + \m_1 \TI_0$.
This potential is very similar to the previous point-mass problem and the equations of motion are simply
\be
\ddvr  = \vF_{01} (\vr) / \beta_{01} \ , \llabel{170911f}
\ee
\be
\dvL_0 = \vT_{01} (\vr) \ , \quad  \dvL_1 = \vT_{10} (\vr)
\ , \llabel{170911g}
\ee
\be
\dvq_0 = \frac{1}{2} (0,\vw_0) \cdot \vq_0 \ , \quad \dvq_1 = \frac{1}{2} (0,\vw_1) \cdot \vq_1 \ , \llabel{170911h}
\ee
where
\begin{eqnarray}
\vF_{kl} (\vr)  \!\!\!\!\! & = & \!\!\!\!\! 
\vf(\m_k, \m_l, \vr) + \vg(\m_k, \TI_l, \vr) + \vg(\m_l, \TI_k, \vr) \nonumber \\ 
 && \!\!\!\!\! + \vh(\m_k, \TI_l, \vr) + \vh(\m_l, \TI_k, \vr)
\ , \llabel{170911f}
\end{eqnarray}
\be
 \vT_{kl} (\vr) =  \vT (\m_l, \TI_k, \vr)
\ , \llabel{170911g}
\ee
$\beta_{kl} = \m_k \m_l / (\m_k + \m_l)$, $\vL_k = \TI_k \vw_k $ is the rotational angular momentum vector of the body with mass $\m_k$, $\vw_k$ its angular velocity and $\vq_k$ the quaternion that represents the rotation from the body frame to the inertial frame.

\subsection{$N$-body problem}
\llabel{nbody}

The previous equations can be easily generalised to the motion of several ellipsoidal bodies.
For a system of $N+1$ bodies, with masses $\m_k$ and inertia tensors $\TI_k$ ($k = 0,1,...,N$), the total potential energy can be written from expression (\ref{170911e}) as
\be 
U = \sum_{k=0}^N \sum_{l>k}^N \left( - \frac{G \m_k \m_l}{\rkl} + \frac{3 G}{2 \rkl^3} \left[ \urkl \cdot \TJ_{kl} \cdot \urkl - \frac{1}{3} \mathrm{tr}(\TJ_{kl}) \right] \right) \ , \llabel{170911i}
\ee
with  $\TJ_{kl} = \m_k \TI_l + \m_l \TI_k$, and $\vrkl = \vR_k - \vR_l$, where $\vR_k$ is the position of the center-of-mass of the body $k$ in the inertial frame.
The equations of motion for each body in the inertial frame are thus
\be
\ddvR_k  = \frac{1}{\m_k} \sum_{l\ne k} \vF_{kl} (\vrkl) \llabel{170911j} \ ,
\ee
\be
\dvL_k = \sum_{l\ne k} \vT_{kl} (\vrkl) \ ,  \llabel{170911k}
\ee
\be
\dvq_k = \frac{1}{2} (0,\vw_k) \cdot \vq_k \ , \llabel{170911l}
\ee
where $\vF_{kl} (\vrkl)$ and $\vT_{kl} (\vrkl)$ are given by expressions (\ref{170911f}) and  (\ref{170911g}), respectively.
As for the two-body problem, we can also express the motion in the relative frame, for instance, with respect to the body with mass $\m_0$.
We let $\vr_k = \vr_{k0} =  \vR_k - \vR_0 \, (k=1,...,N)$.
Thus, 
\be
\ddvr_k  =  \frac{1}{\beta_{0k}} \vF_{0k}(\vr_k) + \sum_{l\ne k} \left[ \frac{1}{\m_k} \vF_{kl} (\vr_k-\vr_l) + \frac{1}{\m_0} \vF_{0l} (\vr_l) \llabel{170911m} \right] \ .
\ee

\section{Conservative dynamics}
\llabel{condyn}

Using the model from section~\ref{nbody}, we now study the dynamics of the \stc system, a hierarchical system of three similar-mass ellipsoidal bodies (see Fig.\,\ref{fig1}).
\bfx{Hierarchical triple stellar systems may also present identical masses, but the \stc system is unique in the small mass regime.}
Unlike stars, whose shape is very close to a sphere, for the small sizes observed in the \stc system (mean radius $\sim 100$~km) we can expect large triaxial asymmetries, which modify the usual point-mass stellar dynamics.
\bfx{In Table\,\ref{tabBen} we list the present best guess on the orbital and physical parameters for the \stc system. The orbital solution is taken from \citet{Benecchi_etal_2010} and the relative sizes from \citet{Mommert_etal_2012}.
All remaining parameters, such as the shape and the spin state, are currently unknown. }

\begin{figure}
\begin{center}
\includegraphics[width=\textwidth]{\figpath 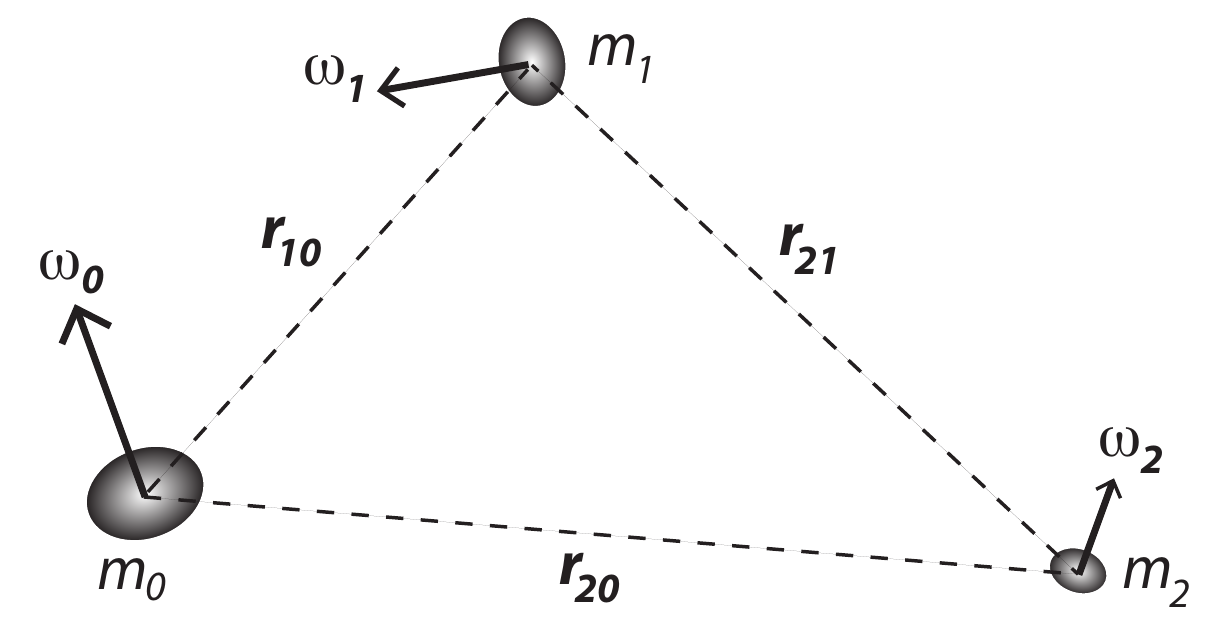} 
 \caption{The \stc system. All bodies are considered oblate ellipsoids. The $\vrkl$ give the relative positions between the center-of-mass of $\m_k$ and $\m_l$, and $\vw_k$ is the rotational angular velocity vector.\llabel{fig1}  }
\end{center}
\end{figure}

\subsection{Shape determination}

The exact shapes and compositions for the three bodies in the \stc system are still unknown.
The integrated spectrum of all components show a very red spectral slope in visible light and a flat spectrum in near infrared \citep{Doressoundiram_etal_2007}. 
The best model reproducing the near infrared spectrum (range 0.37$-$2.33~$\mu$m) includes tholins, serpentine, and crystalline water ice as surface materials \citep{Protopapa_etal_2009}.
The average density of the system is $0.64$~g/cm$^3$, which suggests a high porosity and that these bodies are rubble piles \citep{Mommert_etal_2012}.
Numerical experiments testing the behaviour of cohesionless gravitational aggregates have shown that they evolve into ellipsoidal shapes with pronounced triaxiality \citep{Walsh_etal_2012b}.

We looked in the available data of the solar system for the irregular satellites of the giant planets with similar average radius and densities as those observed for the three components in the \stc system (Table\,\ref{tabBen}). 
The inner binary components {\it Lempo} and {\it Hiisi} (here denoted by bodies 0 and 1, respectively) are almost identical in size and also very similar to Hyperion, which has a mean volumetric radius of 135~km and a mean density of $0.54$~g/cm$^3$ \citep{Thomas_etal_2007}.
We hence assume that the inner bodies have the same semi-axes as Hyperion: $180 \times 133 \times 103$~km.
In turn, the outer circumbinary component {\it Paha} (here denoted by body\,2) is very similar to Epimetheus, which has a mean volumetric radius of 58~km and a mean density of $0.64$~g/cm$^3$ \citep{Thomas_2010}.
We hence assume that the outer body has the same semi-axes as Epimetheus: $65 \times 57 \times 53$~km.

\begin{table}
\caption{Orbital and physical parameters for the (47171) \stc system. \llabel{tabBen}} 
\begin{center}
\begin{tabular}{|l|c|c|c|c|c|c|c|} \hline
\multicolumn{2}{|l|}{fitted parameters$^a$} & \multicolumn{3}{|c|}{\quad orbit 1 \quad \quad} & \multicolumn{3}{|c|}{orbit 2} \\ \hline
orbital period (day)& $P$ & \multicolumn{3}{|c|}{1.9068} & \multicolumn{3}{|c|}{50.302}  \\ 
eccentricity &$e$ & \multicolumn{3}{|c|}{0.101} & \multicolumn{3}{|c|}{0.2949 }  \\ 
inclination (deg) &$i$ & \multicolumn{3}{|c|}{88.9} & \multicolumn{3}{|c|}{79.3 }  \\ 
mean longitude (deg) & $\lambda$ & \multicolumn{3}{|c|}{184.4} & \multicolumn{3}{|c|}{281.1}  \\
longitude of pericenter (deg) & $\varpi$ & \multicolumn{3}{|c|}{47.7} & \multicolumn{3}{|c|}{292.1}  \\
longitude of node (deg) & $\Omega$ & \multicolumn{3}{|c|}{330.0} & \multicolumn{3}{|c|}{325.2}  \\  \hline
\multicolumn{2}{|l|}{estimated parameters} & \multicolumn{2}{|c|}{body\,0} & \multicolumn{2}{|c|}{body\,1} & \multicolumn{2}{|c|}{body\,2} \\ \hline
mass$^b$ $(\times 10^{18} \, \mathrm{kg})$ & $m$  &  \multicolumn{2}{|c|}{6.710} & \multicolumn{2}{|c|}{5.273} & \multicolumn{2}{|c|}{0.767} \\ 
mean radius$^c$ (km) & $R$  & \multicolumn{2}{|c|}{136.0} & \multicolumn{2}{|c|}{125.5} & \multicolumn{2}{|c|}{66.0}  \\ 
Stokes coefficient$^d$ & $J_2$ & \multicolumn{2}{|c|}{0.116} & \multicolumn{2}{|c|}{0.116} & \multicolumn{2}{|c|}{0.050}  \\ 
Stokes coefficient$^d$ & $C_{22}$ & \multicolumn{2}{|c|}{0.029} & \multicolumn{2}{|c|}{0.029} & \multicolumn{2}{|c|}{0.013}  \\ 
Love number$^e$ $(\times 10^{-5}$)& $k_2$ & \multicolumn{2}{|c|}{8.6} & \multicolumn{2}{|c|}{7.2} & \multicolumn{2}{|c|}{1.9}  \\ 
tidal time-lag$^f$ ($\times 10^{2}\,$s) & $\Delta t$ & \multicolumn{2}{|c|}{2.8} & \multicolumn{2}{|c|}{2.8} & \multicolumn{2}{|c|}{68.7}  \\ \hline
\end{tabular}
\end{center}
$^a$ referenced to J2000 equatorial frame at epoch JD\,2453880 \citep{Benecchi_etal_2010};
$^b$ the total mass is $12.75 \times 10^{18}$~kg \citep{Benecchi_etal_2010}, we estimate the relative masses using a uniform density and the mean radius;
$^c$ \citet{Mommert_etal_2012};
$^d$ assuming a triaxiality ratio equivalent to that of Hyperion for bodies 0 and 1 \citep{Thomas_etal_2007}, and that of Epimetheus for body\,2 \citep{Thomas_2010};
$^e$ Eq.(\ref{171013a}) with $\mu = 4$\,GPa;
$^f$ Eq.(\ref{171013b}) with $Q=100$.
\end{table}

Assuming a homogeneous density, we can compute the principal inertia axes ($C \ge B \ge A$) from the ellipsoid semi-axes ($ a \ge b \ge c$)
\be
A = \frac{\m}{5}(b^2+c^2) \ , \quad B = \frac{\m}{5}(a^2+c^2) \ , \quad C = \frac{\m}{5}(a^2+b^2) \ ,
\llabel{171004a}
\ee
as well as the Stokes' gravity field coefficients \citep[e.g.][]{Yoder_1995cnt}, which are the key dynamical parameters:
\be
J_2 = \frac{1}{5} \frac{a^2+b^2-2c^2}{a^2+b^2} \ , \quad \mathrm{and} \quad C_{22} =  \frac{1}{10} \frac{a^2-b^2}{a^2+b^2} \ .
\llabel{171004b}
\ee
The estimated values for the \stc system are given in Table\,\ref{tabBen}.
The real values may differ from these ones, but as it is observed for the remaining irregular satellites \citep[e.g.][]{Thomas_2010}, the order of magnitude is likely correct.
From previous expressions we also deduce that
\be
C_{22} =  \frac{J_2}{2} - \frac{1}{5} \frac{b^2-c^2}{a^2+b^2} 
\quad \Rightarrow \quad 
0 \le C_{22} \le  \frac{J_2}{2} \le 0.1 \ ,
\llabel{171004c}
\ee
so the $J_2$ and $C_{22}$ values must always be in this range.
In our model (section~\ref{model}) we compute the principal axis of inertia from the mean radius and Stokes' coefficients as:
\be
C = \frac{2}{5} \m R^2 \ , \quad \frac{B}{C} = 1 - 5 \left( \frac{J_2}{2} - C_{22} \right)  \ , \quad \frac{A}{C} =  \frac{B}{C} - 10\,C_{22}  \ .
\llabel{171023a}
\ee

\subsection{Present evolution}

\llabel{presentevol}

The orbits of the three objects in the \stc system were determined using Hubble Space Telescope images \citep{Benecchi_etal_2010}.
The best fitted solution, obtained with a Keplerian point-mass model, is provided in Table\,\ref{tabBen}.
We also list the estimated sizes and masses \citep{Mommert_etal_2012}.
The present spin (rotation period and obliquity) of all bodies is unknown.

The rotation period is particularly important for the dynamics of the \stc system.
As for the shape determination, we can try to guess the rotation period value from observations done with other similar objects in the solar system.
The distribution of asteroids and Kuiper-belt objects larger than 200 km in radius show that most rotation periods lie in a narrow 5$-$20~h range irrespective of their other physical properties \citep[e.g.][]{Johansen_Lacerda_2010}.
However, for the tidally evolved objects (such as the Pluto-Charon binary or the satellites of the planets) the rotation period is usually synchronised with the orbital period \citep[e.g.][]{Cheng_etal_2014a}.

In Figure~\ref{fig21} we show three sets of simulations corresponding to different initial rotation periods.
The initial obliquity is always set at $5^\circ$ and the remaining parameters are those listed in Table\,\ref{tabBen}.
For each simulation we show (from top to bottom): the semi-major axis ratio $a/\bar a$ and the eccentricity $e$ of the inner and outer orbit, and the rotation period ratio $P_{\rm rot}/\bar P_{\rm rot}$ and the obliquity $\theta$ of each body in the system.
$\bar a$ and $\bar P_{\rm rot}$ correspond to the average values of the semi-major axis and rotation period, respectively.
We show the relative value in order to better observe the variations of all quantities in the same plot.
We show the evolution for $10^3$~yr (which corresponds to more than $10^5$ inner orbit revolutions), as this short period of time is enough to highlight the main dynamical features of the system.
However, in our simulations we have always integrated over $10^5$~yr.

\begin{figure*}
\begin{center}
\includegraphics[width=\textwidth]{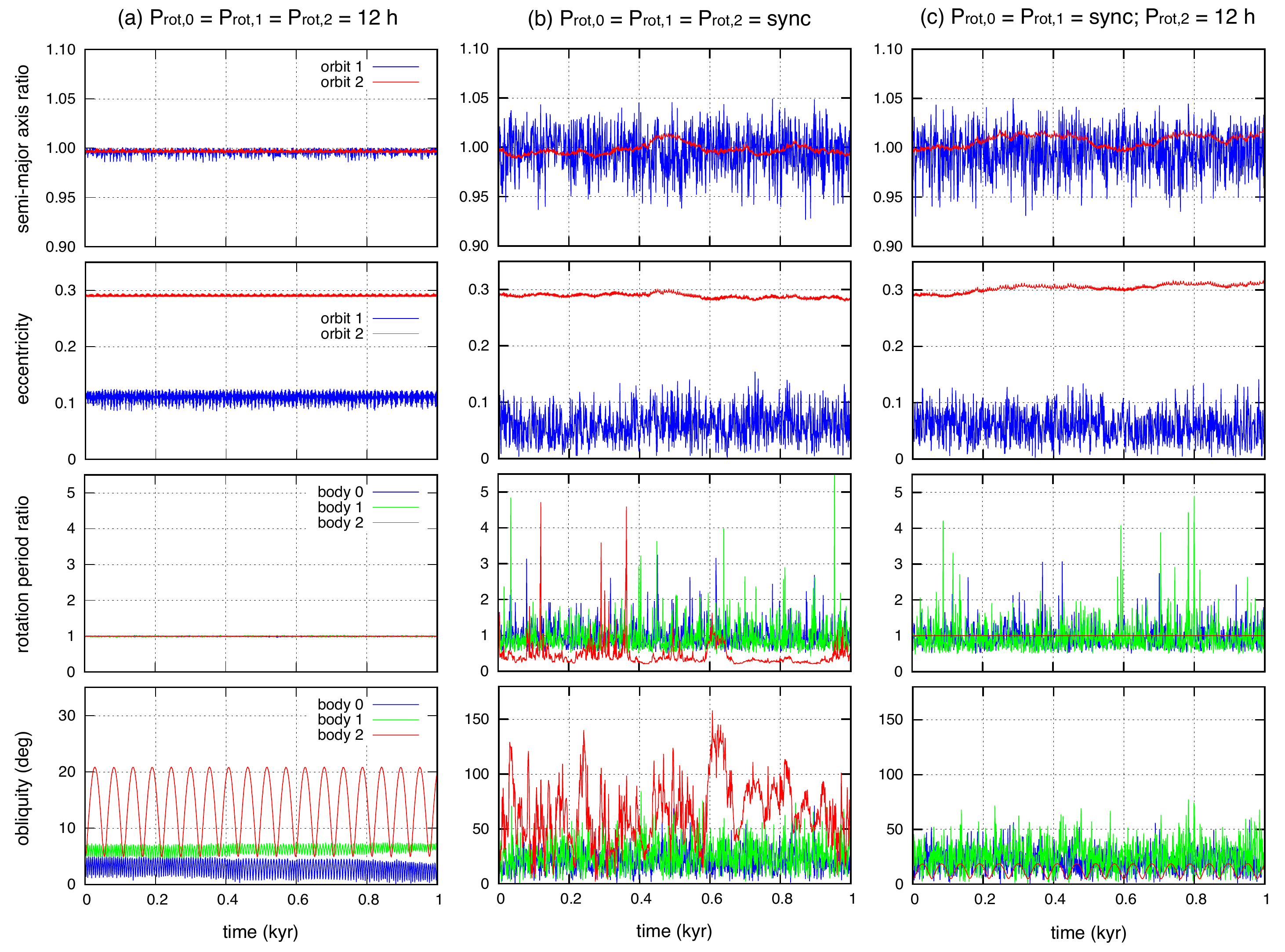} 
 \caption{Orbital and spin evolution of the \stc system (Table\,\ref{tabBen}) over $10^3$~yr for different choices of the initial rotation period: (a) $P_{\rm rot,0} = P_{\rm rot,1} = P_{\rm rot,2} = 0.5$~day; (b) $P_{\rm rot,0} = P_{\rm rot,1} = P_{\rm rot,2} =$ synchronous; (c) $P_{\rm rot,0} = P_{\rm rot,1} = $ synchronous; $P_{\rm rot,2} = 0.5$~day. We show (from top to bottom): the semi-major axis ratio $a/\bar a$ and the eccentricity $e$ of the inner and outer orbit, and the rotation period ratio $P_{\rm rot}/\bar P_{\rm rot}$ and the obliquity $\theta$ of each body in the system.
 \llabel{fig21}  }
\end{center}
\end{figure*}

In the first case (Fig.\ref{fig21}\,a), we set the initial rotation period of the three bodies at 0.5~day (12~h).
At first glance the system appears to be well behaved, both orbital and spin parameters only presenting small variations around their mean values.
However, a more detailed analysis reveals that these variations are not periodic for the orbits and spins of the two inner components.
Indeed, in the obliquity plot we can already see some irregular variations, that become even more pronounced if we extend the simulation for longer periods of time.
By performing a frequency analysis of the orbital mean motion (see section~\ref{famap}) we confirm that moderate chaos exists, only the spin of body\,2 remains regular.

In the second run (Fig.\ref{fig21}\,b), we set the initial rotation period of the three bodies synchronous with the orbital mean motion, i.e., $P_{\rm rot,0} = P_{\rm rot,1} = 1.9079$~day and $P_{\rm rot,2} = 50.053$~day.
Note that these values differ slightly from the orbital periods given in Table~\ref{tabBen}, because we use the exact frequency that is obtained when integrating a point-mass system over $10^3$~yr. 
In this case we observe that both orbits and all three spins are strongly chaotic.
The orbital evolution shown over the initial $10^3$~yr appears to be bounded, but actually both eccentricities can grow to higher values for longer periods of time.
We observe that when the inner orbit eccentricity grows above 0.45, the system becomes unstable (see section~\ref{tidevol}).

In the last case (Fig.\ref{fig21}\,c), we set the initial rotation period of the two inner bodies synchronous with the orbital mean motion and the rotation of the outer body at $P_{\rm rot,2} = 0.5$~day.
This choice is justified by the analysis on the tidal evolution time-scale for this system (section~\ref{evts}).
Tidal dissipation between the two inner bodies is enough to completely despun an initial rotation close to the centrifugal breakup limit.
Therefore, unless the system was formed recently, it is very likely that the two inner bodies are synchronous at present.
On the other hand, tidal dissipation in the outer body is weaker, so the present value may still be very close to the original rotation period.
In this case, we observe that both orbits are still strongly chaotic, as well as the spin of the two inner bodies, only the spin of the outer body remains stable.

\subsection{Simplified system}

\llabel{simpsys}

We observed global chaotic motion for the \stc system.
To better understand the origin of this chaos, here we repeat the same kind of experience as in previous section, but only taking into account the spin of one of the inner components (body\,0).
This choice is supported by the fact that both inner components have similar masses (Table\,\ref{tabBen}) and that we have seen that body\,2 is not the main contributor to the orbital chaos  (Fig.\ref{fig21}\,c).
In Figure~\ref{fig22} we show three sets of simulations corresponding to different shapes and initial rotation periods of body\,0, with the initial obliquity set at $5^\circ$.
The orbital parameters are those listed in Table\,\ref{tabBen}, with bodies\,1 and~2 always taken as point masses.

\begin{figure*}
\begin{center}
\includegraphics[width=\textwidth]{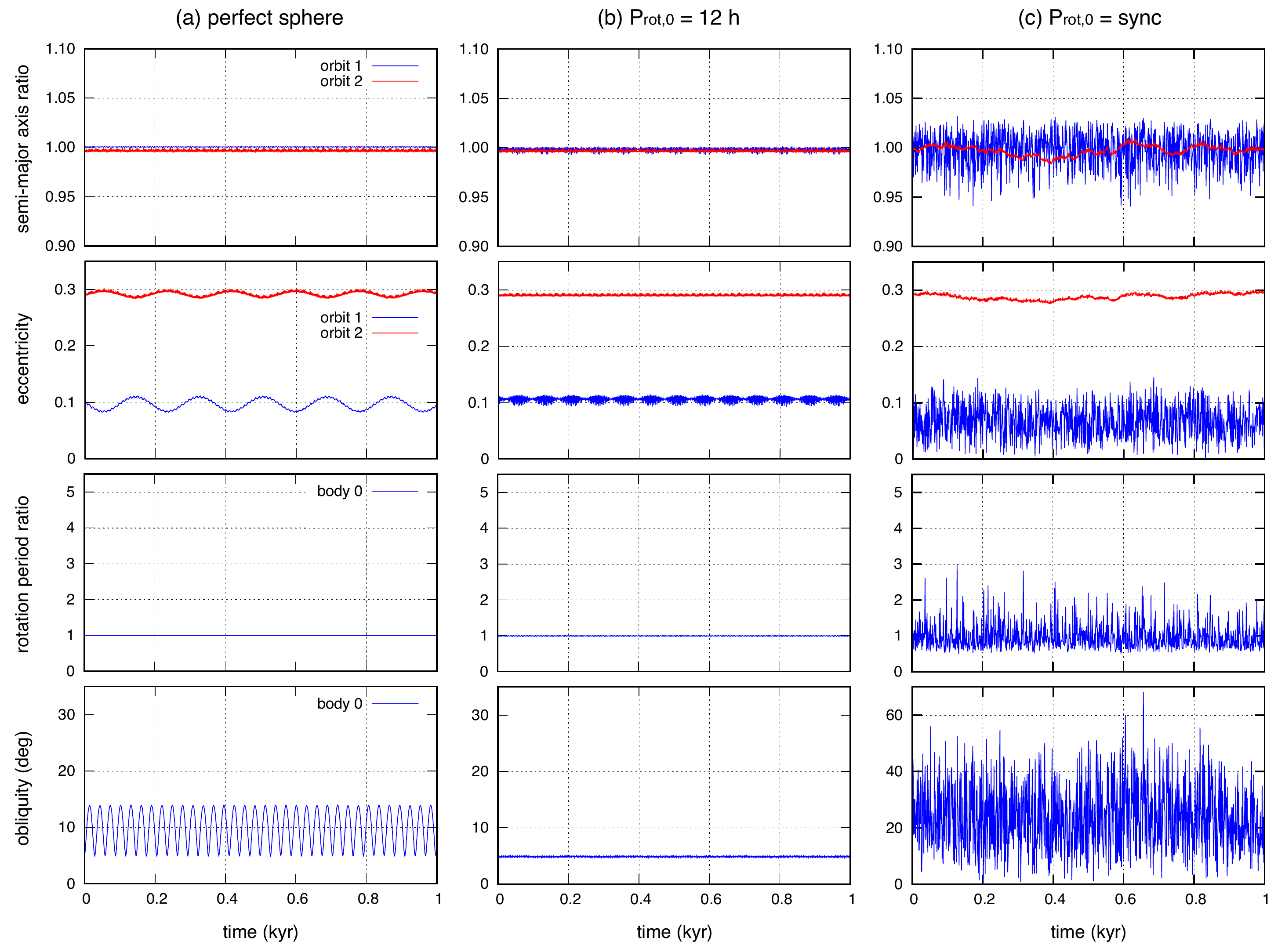} 
 \caption{Orbital and spin evolution of the \stc system (Table\,\ref{tabBen}) over $10^3$~yr for different choices of the shape and initial rotation period of body\,0 (bodies 1 and 2 are point masses): \bfx{(a) perfect sphere (which is equivalent to a point-mass);} (b) $P_{\rm rot,0} = 0.5$~day; (c) $P_{\rm rot,0} =$ synchronous. We show (from top to bottom): the semi-major axis ratio $a/\bar a$ and the eccentricity $e$ of the inner and outer orbit, and the rotation period ratio $P_{\rm rot}/\bar P_{\rm rot}$ and the obliquity.
 \llabel{fig22}  }
\end{center}
\end{figure*}

In the first case (Fig.\ref{fig22}\,a), we set $J_{2,0} = C_{22,0}=0$ for the Stokes' coefficients of body\,0  (which is equivalent to a perfect sphere). 
The initial choice for the rotation period is irrelevant, because in this case body\,0 behaves like a point-mass as well. 
The only difference from the point-mass case is that we are still able to track the evolution of the obliquity (which presents some oscillations due to the precession motion of the orbital plane).
We also follow the evolution of the rotation period, but as expected, it remains unchanged.
The main feature in this case is thus the orbital evolution.
We observe a very regular system with periodic oscillations for both semi-major axes and eccentricities.
The eccentricities are dominated by the secular evolution, while the semi-major axis are almost constant.
In brief, the origin of the orbital chaotic motion must be related to the evolution of the spin(s).

In the second run (Fig.\ref{fig22}\,b), we set $J_{2,0} = 0.116$ and $C_{22,0}=0.029$ (equal to Hyperion's values) and the initial rotation period is $P_{\rm rot,0} = 0.5$~day.
In this case the spin and the orbital motion are still regular.
However, the amplitude of the eccentricity and obliquity cycles is much smaller than before, which means that their values are always very close to the initial ones.
The main reason for this is that the large $J_{2,0} = 0.116$ forces a much faster precession of both spins and orbits.

Finally, in the last case (Fig.\ref{fig22}\,c), we keep $J_{2,0} = 0.116$ and $C_{22,0}=0.029$, but set the initial rotation period at $P_{\rm rot,0} = 1.9079$~day (synchronous).
Contrarily to the previous cases, we now observe a strong chaotic behaviour for both orbits and spin.
The orbital evolution is very similar to the one depicted in Figures~\ref{fig21}(b) and (c), although we now only consider the spin of body\,0. 
We hence conclude that the origin of the orbital chaos is essentially related to the chaotic spin evolution of the inner component(s).
This allow us to simplify the study of the \stc system by restricting it to the study of a single non-spherical body acompained of two point-masses.

\subsection{Stability maps}

\llabel{famap}

The diverse dynamics of the \stc system is mainly controlled by the spin dynamics of one inner body.
The important parameters are the shape and the initial rotation period, which are unknown.
In order to get a full view on the dynamics of the  \stc system it is then important to study the dependence with the shape and the rotation rate.

For that purpose, in this section we draw some stability maps using frequency analysis \citep{Laskar_1990, Laskar_1993PD}.
We adopt the simplified system as reference (section~\ref{simpsys}) and vary the shape and the rotation rate of body\,0.
For each initial condition, we integrate the \stc system over 1000~yr. 
The stability of the orbit is then measured by computing the mean motion $n_1$ and $n_1'$ of the inner orbit over two consecutive time intervals of 500~yr.
The difference $D=|1-n_1'/n_1|$ is a measure of the chaotic diffusion of the trajectory.
It should be very close to zero for regular motion and $\sim 1$ for strong chaotic motion \citep[for more details see][]{Robutel_Laskar_2001, Correia_etal_2005, Couetdic_etal_2010}.
In the present case, regular motion will require $D \lesssim 10^{-6}$ (blue colors), solutions with $10^{-6} \lesssim D \lesssim 10^{-5}$ (green-yellow colors) correspond to chaotic rotation, but still nearly regular orbits (bounded and marginal chaos for the orbits), and $D \gtrsim 10^{-5}$ (orange-red colors) corresponds to strong chaotic motion for the orbits and spins. 
For $D \gtrsim 10^{-3}$ the system is unstable in a short period of time (sometimes shorter than the time-span of the integrations).

In Figure~\ref{figj2} we study the impact of the shape given by different $J_{2,0} $ and $C_{22,0}$ values, from an almost perfect sphere ($10^{-10}$) until the maximal allowed values (Eq.\,(\ref{171004c})).
The initial rotation of body\,0 is synchronous ($P_{\rm rot,0} = 1.9079$~day) and the initial obliquity is $5^\circ$.
The other two bodies are point-masses and the initial conditions for the orbits are the present ones (Table\,\ref{tabBen}).
We observe that the critical parameter for stability is the $C_{22}$ value.
For $C_{22,0}<10^{-7}$ and $J_{2,0}<10^{-5}$ the system is completely stable. 
The $J_2$ appears to have some impact in the stability only in the range $10^{-5}<J_{2,0}<10^{-3}$, but it is essentially marginal chaos, the system is not destroyed. 
For  $10^{-7}<C_{22,0}<10^{-5}$ marginal chaos also appears for $J_{2,0}<10^{-5}$, but interestingly the system remains very stable for $J_{2,0}>10^{-3}$.
The large $J_2$ values render the system more stable likely because the precession rate of the spin and orbits increases with $J_2$, which helps to cancel the effect from the $C_{22}$.
Finally, for $C_{22,0}>10^{-5}$ the system is always unstable.
This result is quite remarkable, because $C_{22,0}\sim10^{-5}$ is a typical value observed for planets like Mercury and Mars \citep{Yoder_1995cnt}, which are almost spherical.
For comparison, a black dot marks the position of the solution displayed in Fig.\,\ref{fig22}\,c, which corresponds to our best guess of the $C_{22}$ values for the \stc bodies.
Therefore, unless the two inner bodies in the \stc system are atypically spherical, the present best fitted orbital solution \citep{Benecchi_etal_2010} is chaotic and unstable for near synchronous rotation.

\begin{figure}
\begin{center}
\includegraphics[width=\columnwidth]{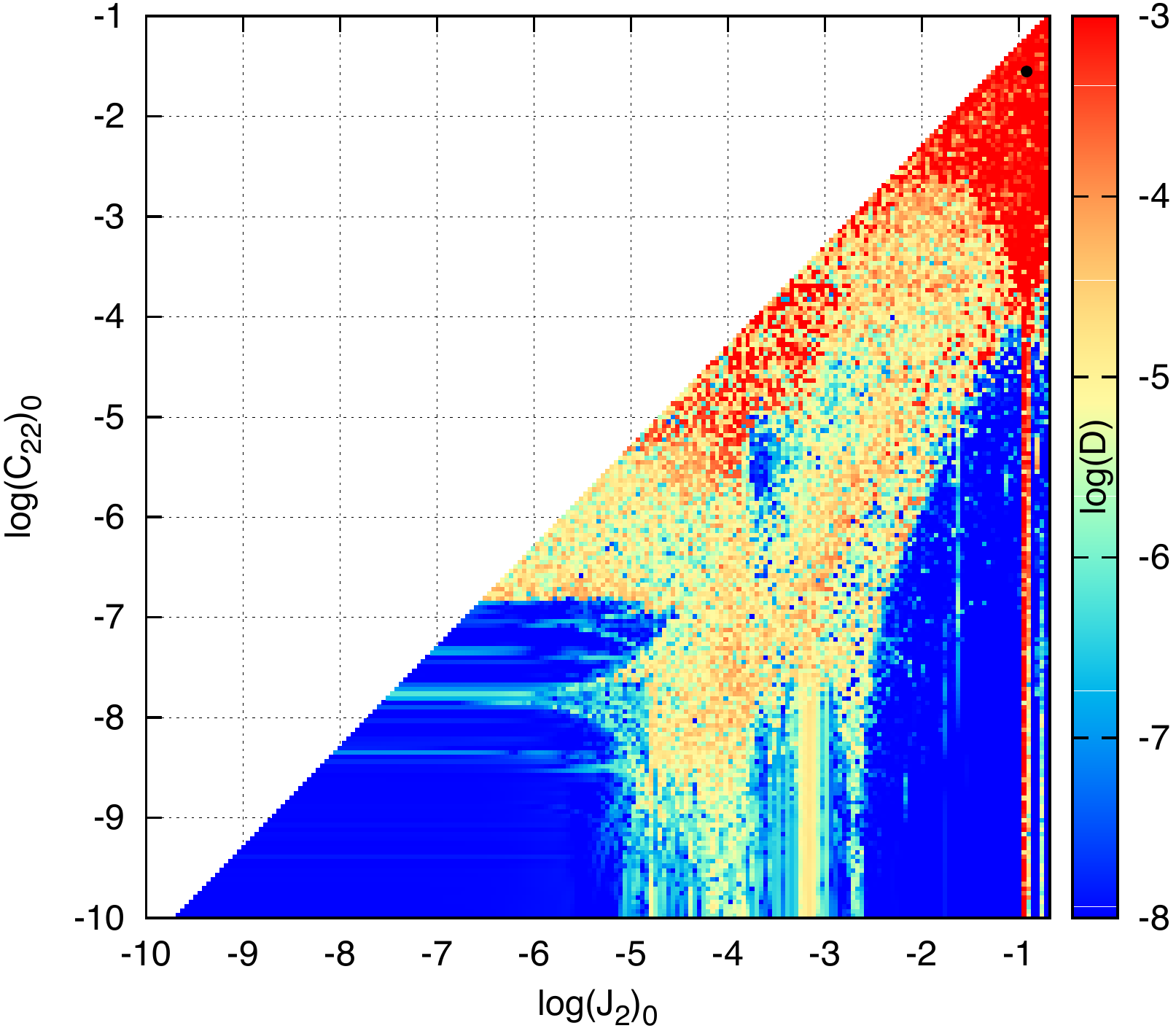} 
 \caption{Stability analysis of the \stc system for different shapes of the body\,0.
The $J_{2,0} $ and $C_{22,0}$ values are varied in a logarithmic scale with step size $0.05$, from an almost perfect sphere ($10^{-10}$) until the maximal allowed values (Eq.\,(\ref{171004c})).
The white zone corresponds to impossible combinations of the pair ($J_{2,0}, C_{22,0}$) according to expression (\ref{171004c}).
The initial rotation of body\,0 is synchronous ($P_{\rm rot,0} = 1.9079$~day) and the initial obliquity is $5^\circ$.
The other two bodies are point-masses and the initial conditions for the orbits are the present ones (Table\,\ref{tabBen}).
For comparison, a black dot marks the position of the solution displayed in Fig.\,\ref{fig22}\,c.
The color index $D$ is explained in the second paragraph of section~\ref{famap}.
 \llabel{figj2}  }
\end{center}
\end{figure}

Since the $J_2$ coefficient is not a critical shape parameter for stability studies, henceforward we restrict the impact of the shape on the dynamics to the analysis of the $C_{22}$ coefficient.
However, according to expression (\ref{171004c}) the two Stokes coefficients are related: by increasing $C_{22}$ we also increase $J_2$.
Thus, to simplify our analysis, for each $C_{22}$ we will assume the minimal corresponding value for $J_2$, that is, $J_2 = 2 C_{22}$.
This choice actually corresponds to take $J_2$ values along the diagonal straight line that divides the stability map from Fig.\,\ref{figj2} in possible and impossible initial conditions. 
As we can see, this border line roughly corresponds to an upper limit for chaotic solutions with a given $C_{22}$ value.

In previous sections we saw that a fast rotation rate may also help to stabilise the system.
Therefore, in Figure~\ref{figw0} we study the stability of the system for different $C_{22,0}$ and for the initial rotation rate of body\,0, $|\vw_0| = 2\pi / P_{\rm rot,0} $ ranging from $-0.6\,n_1$ (retrograde) until $3.1\,n_1$.
The initial obliquity is $5^\circ$, the other two bodies are point-masses, and the initial conditions for the orbits are the present ones (Table\,\ref{tabBen}).
We observe that for $C_{22,0} > 10^{-2}$ there is a large chaotic zone between $n_1/2$ and $5n_1/2$.
The origin of this chaos corresponds to the overlap of the individual spin-orbit resonances libration widths \citep{Wisdom_etal_1984, Wisdom_1987}.
We stress, however, that the chaos measured in Fig.\,\ref{figw0} is the orbital chaos, that is, the spin is chaotic, but the orbits are also very chaotic, corresponding to a system that is not stable.
As we decrease the $C_{22,0}$ value, the chaotic areas around each spin-orbit resonance shrink, but they persist in a tiny region surrounding the synchronous resonance up to $C_{22,0} > 10^{-7}$.
This explains why in Fig.\,\ref{figj2} we observed strong chaos for $C_{22,0} > 10^{-5}$ and marginal chaos for  $10^{-5}>C_{22,0}>10^{-7}$.

The orbital period of the inner orbit is $\sim 1.9$~days (Table\,\ref{tabBen}), so the maximal rotation rate of $3.1\,n_1$ shown in Fig.\,\ref{figw0} corresponds to about $14.7$~h, which is close to the 12~h example shown in Fig.\,\ref{fig22}.
In this region we observe that stability is already possible for all $C_{22,0}$ values, provided that we are not close to a spin-orbit resonance.
Indeed, none of these resonances is stable for large $C_{22,0}$. 
Therefore, for a body decreasing its rotation rate under the action of tidal dissipation, the orbits certainly become chaotic when the spin arrives in the near synchronous regime.
We conclude that unless tides are exceptionally weak, the present best fitted orbital solution \citep{Benecchi_etal_2010} is chaotic and unstable for large $C_{22,0}$ values.

\begin{sidewaysfigure}
\includegraphics[width=\textwidth]{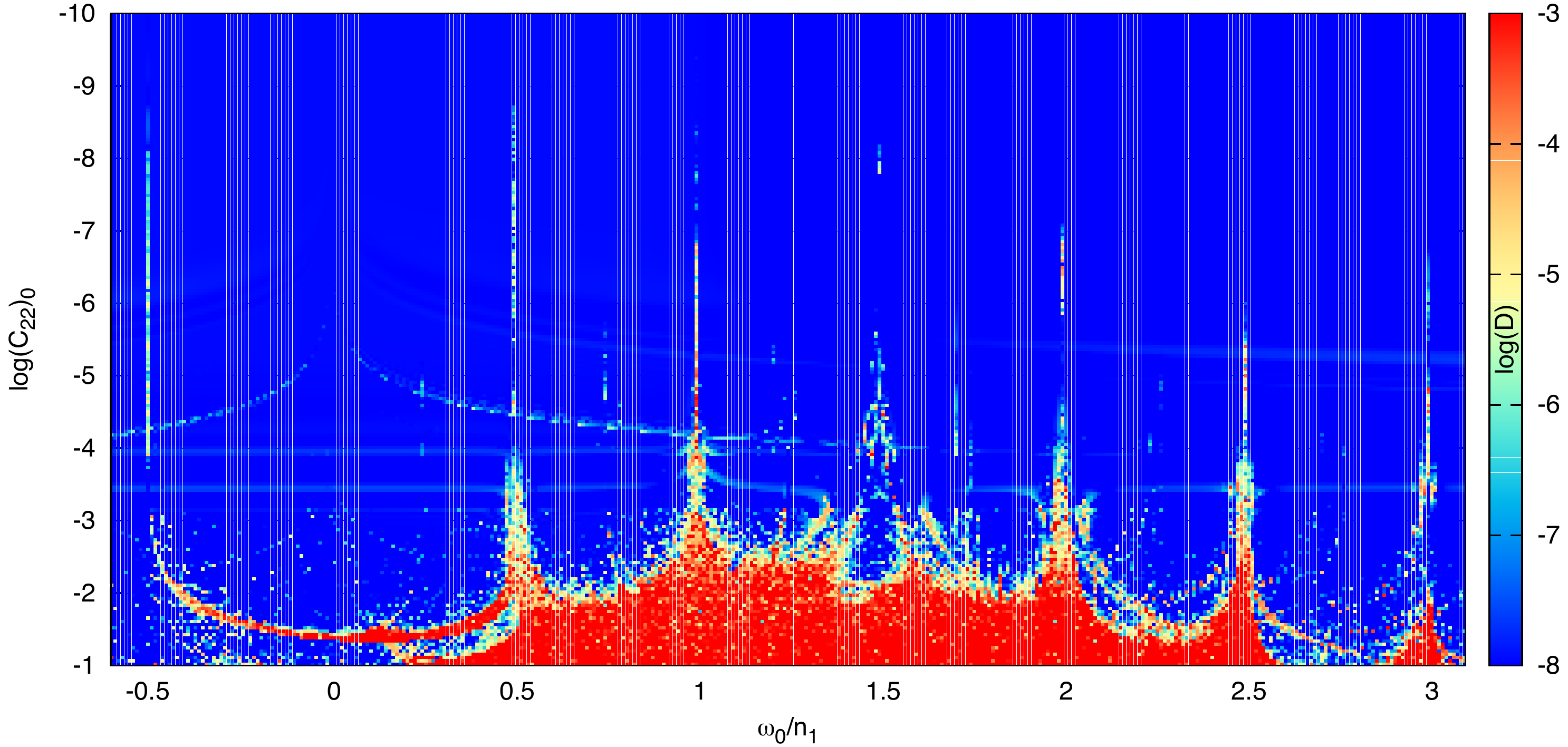} 
 \caption{Stability analysis of the \stc system for different shapes and initial rotations of the body\,0.
The $C_{22,0}$ values are varied in a logarithmic scale with step size $0.05$, from an almost perfect sphere ($10^{-10}$) until its maximal value, and $J_{2,0} = 2 C_{22,0}$ (Eq.\,(\ref{171004c})).
The initial rotation rate $|\vw_0| / n_1 $ is varied linearly with step size 0.01 from $-0.6$ (retrograde) until $3.1$.
The initial obliquity is $5^\circ$, the other two bodies are point-masses and the initial conditions for the orbits are the present ones (Table\,\ref{tabBen}).
The color index $D$ is explained in the second paragraph of section~\ref{famap}.
 \llabel{figw0}  }
\end{sidewaysfigure}

\subsection{Initial orbital solution}

\llabel{initorb}

One important reason why chaotic and unstable orbits exist in the \stc system is because the present eccentricity of the inner orbit is relatively high ($\sim 0.1$). 
Non-zero eccentricity introduces multiple spin-orbit resonances that overlap for large $C_{22}$ values.
For zero eccentricity, only the synchronous resonance persists, so we could expect a more regular system.

The best-fitted orbital solution listed in Table\,\ref{tabBen} was obtained using simple non-interacting Keplerian orbits \citep{Benecchi_etal_2010}.
However, we have seen that the eccentricity undergoes rapid and chaotic variations in a short period of time (Figs.\,\ref{fig21} and \ref{fig22}) that cannot be reproduced with a simple Keplerian model.
Even when we exclude the impact of the spins in the orbits (Fig.\,\ref{fig22}\,a), a full eccentricity cycle only takes about 100~yr. 
Therefore, over the course of the already existing observations (7~yr) we should be already fitting the observational data with a more realistic model like the one presented in section~\ref{nbody}.
It is then possible that the present eccentricity is incorrectly determined.

In Figure~\ref{fige0} we study the stability of the system for different $C_{22,0}$ and initial eccentricities of the inner orbit ranging from 0 until 0.99.
The initial rotation of body\,0 is synchronous ($P_{\rm rot,0} = 1.9079$~day) and the initial obliquity is $5^\circ$.
The other two bodies are point-masses and the initial conditions for the orbits are the present ones (Table\,\ref{tabBen}), except for the inner orbit eccentricity.

\begin{figure}
\begin{center}
\includegraphics[width=\columnwidth]{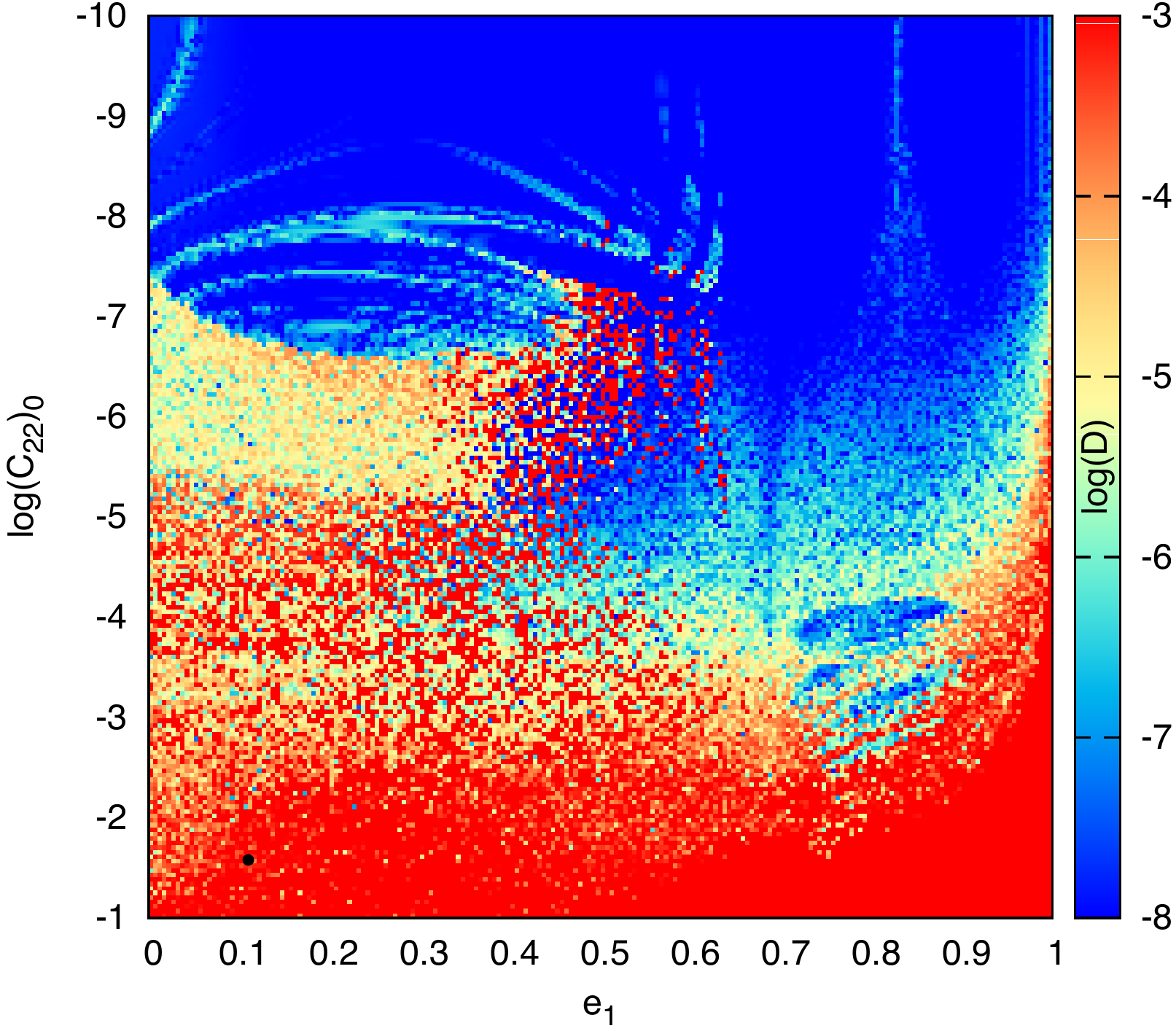} 
 \caption{Stability analysis of the \stc system for different shapes of the body\,0 and initial eccentricities of the inner orbit.
The $C_{22,0}$ values are varied in a logarithmic scale with step size $0.05$, from an almost perfect sphere ($10^{-10}$) until its maximal value, and $J_{2,0} = 2 C_{22,0}$ (Eq.\,(\ref{171004c})).
The initial eccentricity of the inner orbit $e_1$ is varied linearly with step size 0.005 from 0 until 0.995.
The initial rotation of body\,0 is synchronous and the initial obliquity is $5^\circ$.
The other two bodies are point-masses and the initial conditions for the orbits are the present ones (Table\,\ref{tabBen}), except for the inner orbit eccentricity.
For comparison, a black dot marks the position of the solution displayed in Fig.\,\ref{fig22}\,c.
The color index $D$ is explained in the second paragraph of section~\ref{famap}.
 \llabel{fige0}  }
\end{center}
\end{figure}

We observe that for $C_{22,0} < 10^{-7}$ the system is always stable, even for very high values of the eccentricity. 
Note that stable does not mean that the eccentricity is constant, only that it varies regularly (see Fig.\,\ref{fig22}\,a).
Surprisingly, for $10^{-7} < C_{22,0} < 10^{-5}$ the system remains stable for initial eccentricities higher than $\sim 0.5$, but marginal chaos sets in for smaller values.
More interestingly, for $10^{-5} < C_{22,0} < 10^{-4}$ the system is still stable for initial eccentricities in the range $0.5 < e_1 < 0.9$, but strongly chaotic outside this interval.
This behaviour persists for $10^{-4} < C_{22,0} < 10^{-3}$ in the range $0.7 < e_1 < 0.9$.
For $C_{22,0} > 10^{-3}$ the orbits become strongly chaotic and unstable for any initial eccentricity value, and for $e_1 > 0.5$ the systems is even destabilised before $10^3$~yr (the time interval that we are using to measure the chaotic indicator).

The analysis of the stability for different values of the initial eccentricity of the inner orbit allows us to draw two important conclusions.
First, for very high values of the initial eccentricity the system is more stable.
A possible reason is that for high eccentricities the resonance width of individual resonances is much larger than their mutual separation, such that they completely overlap and encompass the synchronous resonance \citep[e.g.][]{Morbidelli_2002}.
Second, for $0 < e_1 < 0.4$ all orbits have a similar behavior to the case with initial $e_1=0.101$ (Table\,\ref{tabBen}).
This means that even initial circular orbits experience strong chaotic motion for $C_{22,0} > 10^{-5}$, similar to the exemples shown in Figures~\ref{fig21} and \ref{fig22}.
This result could be somehow anticipated, since the chaotic eccentricity variations with initial $e_1=0.101$ can assume values very close to zero at some point in the evolution.

In the next section we show that circular orbits are nevertheless possible for high $C_{22}$ values after some tidal evolution.
However, this requires a simultaneous adjustment of all initial parameters, it is not enough to only set the initial eccentricity of the inner orbit at zero.
The orbital evolution of a 3-body system is coupled, tidal dissipation in the inner orbit is communicated to the outer one in a way that the final equilibrium values for the eccentricity and pericenter position of the inner orbit depend on the properties of the outer orbit \citep{Wu_Goldreich_2002, Mardling_2007, Laskar_etal_2012}.
Therefore, for a given outer orbit, we need to search for stable eccentricity configurations by also exploring the longitude of the pericenter.

In Figure~\ref{figew} we study the stability of the \stc system for the best estimation of the $J_2$ and $C_{22}$ values (Table\,\ref{tabBen}), but for different initial eccentricities and pericenter values of the inner orbit.
In a first experiment (Fig.\,\ref{figew}\,a) we only consider the spin of body~0 as in Fig.\,\ref{fige0}, i.e., we adopt the simplified case where bodies~1 and~2 are point-masses (as in sections~\ref{simpsys} and~\ref{famap}). 
In a second experiment (Fig.\,\ref{figew}\,b), the spin of all bodies is considered (as in section~\ref{presentevol}).
The initial rotation period of the two inner bodies is synchronous, the rotation period of the outer body is 12~h, the initial obliquities are $5^\circ$, and the remaining orbital parameters are the present ones (Table\,\ref{tabBen}). 
We observe that, in spite of the large $C_{22}$ values, two stability islands exist for small eccentricities around $\varpi = 130^\circ$ and $\varpi = 310^\circ$.
In the simplified case (Fig.\,\ref{figew}\,a) these islands are large and stability is possible for initial eccentricities as high as $e_1 = 0.07$, which is close to the observed value $e_1 = 0.101 \pm 0.006$ \citep{Benecchi_etal_2010}.
In the full case (Fig.\,\ref{figew}\,b), the stability areas are smaller and the maximal stable eccentricity decreases to $e_1=0.03$.
Although stability is possible in some regions with non-zero eccentricity, the observed longitude of the pericenter $\varpi_1 = 47.7^\circ \pm 6.3^\circ$ \citep{Benecchi_etal_2010} is very distant from the stable areas. 
For a better comparison, we show the best fitted solution (Table\,\ref{tabBen}) with a black dot.
Note, however, that the stable islands also depend on the determination of the outer orbit. 
Usually, astrometric measurements provide better constrains for the outer orbit, but its determination is also subject to some uncertainties. 

\begin{figure}
\begin{center}
\includegraphics[width=\columnwidth]{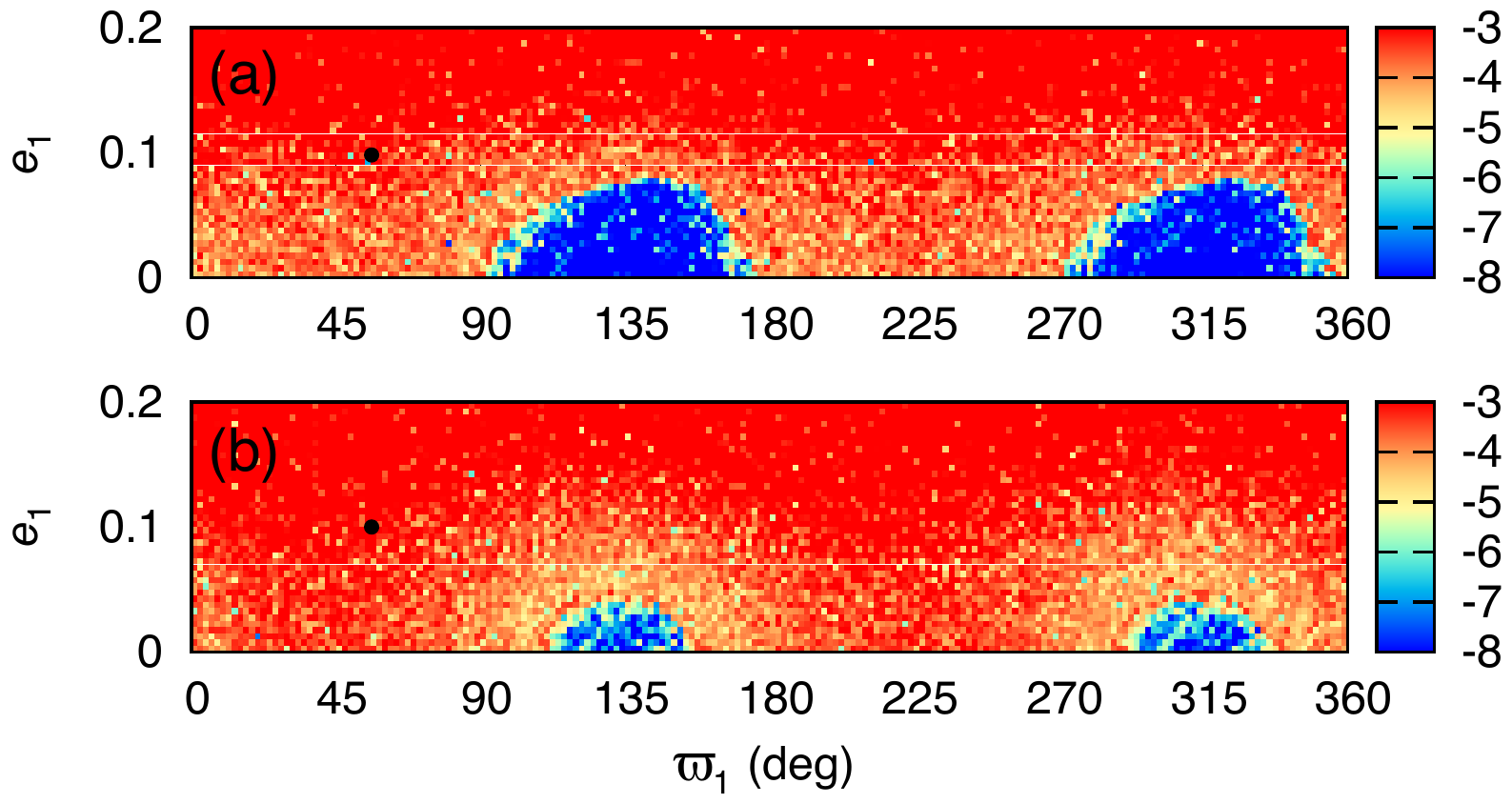} 
 \caption{Stability analysis of the \stc system for different initial eccentricities and longitudes of the pericenter of the inner orbit.
The initial eccentricity $e_1$ is varied linearly with step size 0.005 from 0 until 0.20, while the longitude of the pericenter $\varpi_1$ is varied linearly with step size $1.8^\circ$ from $0^\circ$ until $360^\circ$.
In (a) we only consider the spin of body~0, bodies~1 and~2 are point-masses. 
In (b), the spin of all bodies is considered.
The $J_2$ and $C_{22}$ values correspond to our the best estimation (Table\,\ref{tabBen}), the initial rotation  of the two inner bodies is synchronous, the rotation of the outer body is 12~h, the initial obliquities are $5^\circ$, and the remaining orbital parameters are the present ones (Table\,\ref{tabBen}).
For comparison, a black dot in (a) marks the position of the solution displayed in Fig.\,\ref{fig22}\,c, and in (b) the solution displayed in Fig.\,\ref{fig21}\,c.
The color index $D$ is explained in the second paragraph of section~\ref{famap}.
 \llabel{figew}  }
\end{center}
\end{figure}

\section{Tidal evolution}

\llabel{tidevol}

In previous sections we have seen that the rotation period \bfx{of the inner bodies} is an important variable regarding the stability of the \stc system.
In particular, it is important to determine if the present rotation can still be close to the primordial (fast) rotation, or if it is nearly synchronous due to tidal dissipation.

\subsection{Tidal dissipation for small bodies}

Tidal dissipation is usually modelled through the second Love number $k_2$ and the quality factor $Q$.
The first is related to the rigidity of the body and measures the amplitude of the tidal deformation, while the second is related with the viscosity and measures the amount of energy dissipated in a tidal cycle \citep[e.g.][]{Munk_MacDonald_1960}.

The Love number for an incompressible homogeneous elastic body is given by \citep{Love_1911}
\be
k_{2} = \frac{3}{2}\left(1+\frac{19\mu}{2g\rho R}\right)^{-1} \ ,
\llabel{171013a}
\ee
where $\rho$ is the mean density, $g = G m / R^2$ is the surface gravity, and $\mu$ is the rigidity.
The temperature of the \stc system is about 40\,K and its low density $\sim 0.6$~g/cm$^3$ suggests porous water ice bodies \citep[e.g.][]{Protopapa_etal_2009}.
It is common to estimate $\mu \approx 4$\,GPa for icy bodies \citep[e.g.][]{Nimmo_Schenk_2006}, so we obtain from previous expression $k_2 \approx 10^{-4}$ for the two inner bodies and $k_2 \approx 10^{-5}$ for the outer body (see Table\,\ref{tabBen}).

The quality factor $Q$ depends on the viscosity of the body, but also on the tidal model that we adopt for the dissipation. 
In addition, measurements on the $Q$ factor are usually indirect and difficult to obtain. 
The best measurements are for the Earth and Mars, for which we have $Q \sim 10$ \citep{Dickey_etal_1994} and $Q \sim 80$ \citep{Lainey_etal_2007}, respectively.
However, in the case of the Earth, the present value is dominated by the oceans, the Earth's solid body $Q$ factor is estimated to be 280 \citep{Ray_etal_2001}.
Indirect $Q$ values derived from the orbital evolution of some satellites also find that $Q<500$ \citep{Goldreich_Soter_1966}.
Therefore, it is commonly accepted that for solid bodies $Q\sim100$, so we also adopt this value in the present work.

\subsection{Evolution time-scale}

\llabel{evts}

Several possible models for the formation of asteroid and Kuiper-belt binaries have been proposed, including fission, dynamical capture, and collision \citep[for a review see][]{Noll_etal_2008}.
For the \stc triple system we can have different origines for each component, for instance, fission or collision for the inner pair and dynamical capture for the outer one.
Fission requires that the spin rotates faster than the centrifugal breakup period $\sqrt{3\pi/G\rho}$, which gives about 4~h for the \stc bodies.
We can thus assume this rotation period as the maximal value for the initial rotation just after formation.

The spin-down time $t_{sd}$, i.e., the time that a body takes to reach the near synchronous rotation, can be estimated through \citep[e.g.][]{Peale_1977, Gladman_etal_1996}
\be
t_{sd} \approx   \frac{C |\vw_0| a_0}{3 G \M^2} \left(\frac{a_0}{R}\right)^5 \frac{Q}{k_2} \ ,
\llabel{171016z}
\ee
where $a_0$ is the initial semi-major axis, $|\vw_0|$ is the initial rotation rate, and $\M$ is the mass of the perturber.
Using the centrifugal breakup period in previous expression gives
\be
t_{sd} \approx   \frac{P_0}{15 \pi}  \left(\frac{\m+\M}{\M}\right)^{1/2} \left(\frac{\m}{\M}\right)^{3/2} \left(\frac{a_0}{R}\right)^{9/2} \frac{Q}{k_2} \ ,
\llabel{171017a}
\ee
where $P_0$ is the initial orbital period.
Adopting the best estimated values listed in Table\,\ref{tabBen} we get $t_{sd} \approx 1$~Myr for the inner binary and $t_{sd} \approx 1$~Gyr for the outer body.
We hence conclude that, unless the system was formed very recently (only a few Myr ago), the rotation of the two inner bodies is most likely in the near-synchronous regime, while the rotation of the outer body can still be close to its primordial value.

\subsection{Tidal model}

The equations of motion derived in section~\ref{secmodel} are only valid in the conservative case (without tidal dissipation).
In order to take into account tidal evolution we need to adopt a tidal model and complete these equations.
A large variety of tidal models exist, the most commonly used are the constant$-Q$ \citep[e.g.][]{Munk_MacDonald_1960}, the linear model \citep[e.g.][]{Mignard_1979}, 
and the Maxwell model \citep[e.g.][]{Correia_etal_2014}.
Some models appear to be best suited for some situations than others, but there is no model that is globally accepted.
Anyway, whatever is the tidal model adopted, the qualitative conclusions are more or less unaffected, the system always evolves into an energy minima \citep[e.g.][]{Hut_1980, Adams_Bloch_2015}.

The dissipation of the mechanical energy of tides in the body's interior is responsible for a time delay $\Delta t$ between the initial perturbation and the maximal deformation.
For a given tidal frequency, $\sigma$, the tidal dissipation can be related to this delay through \citep[e.g.][]{Correia_Laskar_2003JGR}
\be
Q_\sigma^{-1} = \sin ( \sigma \Delta t_\sigma ) \approx \sigma \Delta t_\sigma \ .
\llabel{171013b}
\ee
The exact dependence of $\Delta t_\sigma$ on the tidal frequency is unknown.
As in previous studies with minor bodies \citep[e.g.][]{Cheng_etal_2014a, Correia_etal_2015, Quillen_etal_2017}, we adopt here a viscous linear model for tides \citep{Singer_1968, Mignard_1979}.
In this model it is assumed that the time delay is constant and independent of the frequency.
Since we adopt $Q=100$, we also need to fix the tidal frequency in order to get a constant value for $\Delta t$.
For simplicity, we set $\sigma = n$, which gives $\Delta t \approx 10^2$~s for the two inner bodies and $\Delta t \approx 10^3$~s for the outer one (see Table\,\ref{tabBen}).

The linear tidal model provides very simple expressions for the tidal interactions, valid for any eccentricity, inclination, rotation and obliquity.
As in section~\ref{pmp}, consider an ellipsoidal body with mass $\m$ that orbits a point-mass $\M$ located at $\vr$.
The total tidal force acting on the orbit is given by \citep[e.g.][]{Mignard_1979}
\be
\vF_t (K, \M, \vw, \vr) = - K \frac{\M^2}{r^{10}} \Big[ 2 (\vr \cdot \dvr ) \vr + r^2 (\vr \times \vw + \dvr ) \Big] \ , 
\llabel{171016a}
\ee
and the tidal torque on the spin
\be
\vT_t (K, \M, \vw, \vr) = - \vr \times \vF_t = K \frac{\M^2}{r^8} \Big[ (\vr \cdot \vw) \vr - r^2\vw + \vr \times \dvr \Big]  \ ,
\llabel{171016c}
\ee
where
\be
K =  3 k_2 G R^5 \Delta t
\llabel{171016b}
\ee
contains all the quantities pertaining to the body with mass $\m$.
We can now rewrite equations (\ref{170911j}) and (\ref{170911k}) to take into account the tidal evolution of the orbits and spins of a $N$-body system as
\be
\ddvR_k  = \frac{1}{\m_k} \sum_{l\ne k} \Big[ \vF_{kl} (\vrkl) + \vF_t (K_k, \m_l, \vw_k, \vrkl) + \vF_t (K_l, m_k, \vw_l, \vrkl) \Big]
\ , \llabel{170911jc} 
\ee
\be
\dvL_k = \sum_{l\ne k} \Big[ \vT_{kl} (\vrkl) + \vT_t (K_k, \m_l, \vw_k, \vrkl) \Big] 
\ .  \llabel{170911kc}
\ee

\subsection{Numerical simulations}

In a first experiment we run a simulation with and without tidal effects for the present best guess for the \stc system parameters (Table\,\ref{tabBen}).
The initial rotation period of the two inner bodies is synchronous, the rotation of the outer body is 12~h, and the initial obliquities are $5^\circ$.
This example is exactly the same as the one used for the example shown in Fig.\,\ref{fig21}\,(c), described in section~\ref{presentevol}.
In Figure\,\ref{tides1} we show the evolution of the semi-major axes and eccentricities.

\begin{figure*}
\begin{center}
\includegraphics[width=\textwidth]{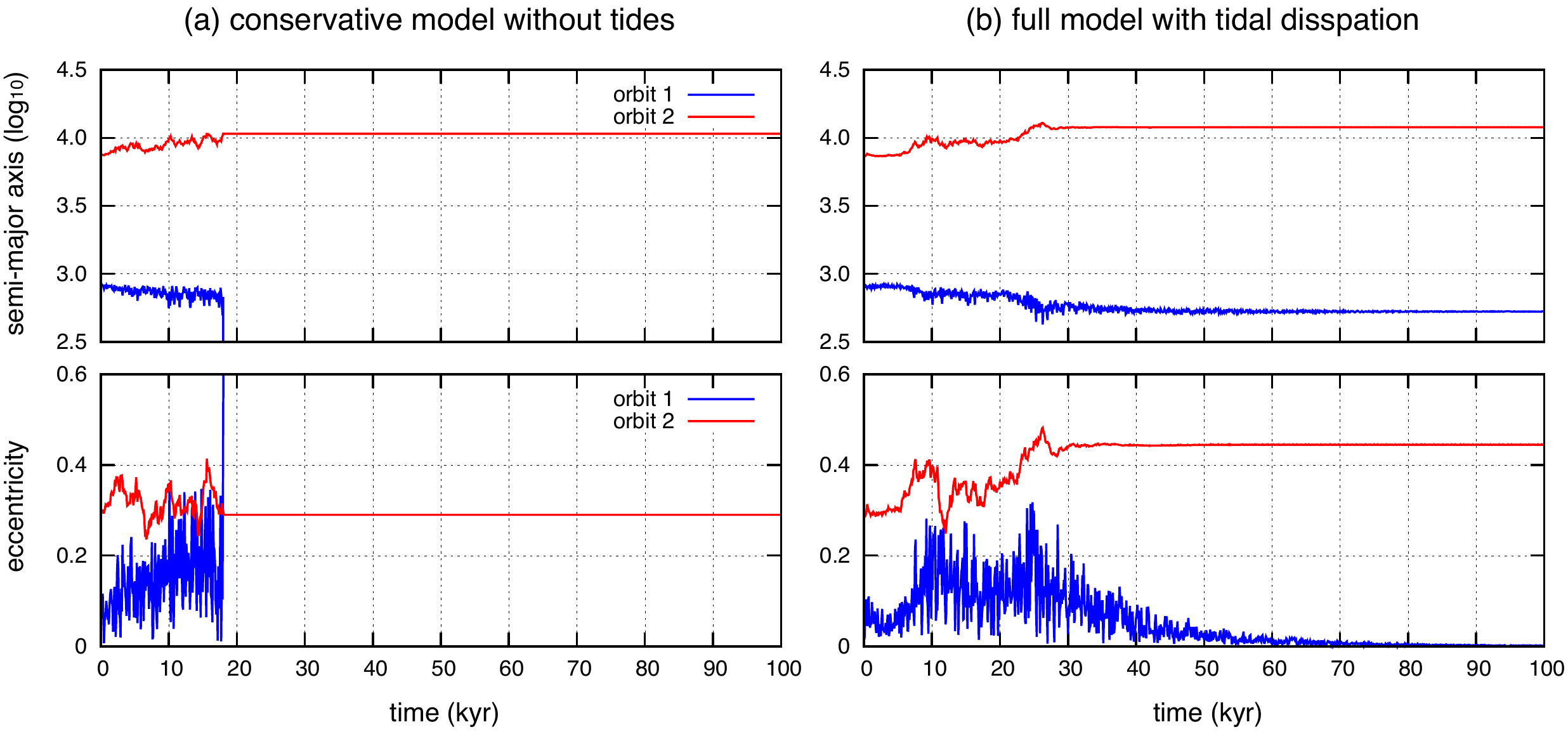} 
 \caption{Orbital evolution of the \stc system (Table\,\ref{tabBen}) over $10^5$~yr with (a) and without (b) tidal effects.
The initial rotation period of the two inner bodies is synchronous, the rotation of the outer body is 12~h, and the initial obliquities are $5^\circ$. In the top panel we show the semi-major axes $\log_{10} a_{[\mathrm{km}]}$, while in the bottom panel we show the eccentricities. We observe that the inclusion of tides is able to stabilise the system.
 \llabel{tides1}  }
\end{center}
\end{figure*}

When tides are not included (Fig.\,\ref{tides1}\,a), we observe that the system is very chaotic, and it is destroyed after 18~kyr.
We also run additional 100 simulations with slightly different initial eccentricities for the inner orbit ($e_1 \in [0.100, 0.101]$), and the system always becomes unstable before 30~kyr.
Instability is triggered when the inner orbit eccentricity becomes higher than 0.45 or when the outer orbit eccentricity becomes higher than 0.9.

When tides are included (Fig.\,\ref{tides1}\,b), we observe that the system is still very chaotic, but after some wandering in the chaotic zone the eccentricity of the inner orbit is damped. 
After this stage the system is no longer chaotic.
We also run additional 100 simulations with slightly different initial eccentricities for the inner orbit ($e_1 \in [0.100, 0.101]$). In 42 cases the system is circularised, while in the remaining 58 the system is still unstable.
The systems that are destroyed correspond to those for which the inner orbit eccentricity increases very fast and overcomes the 0.45 threshold before tides manage to efficiently damp it.

We can conclude that, in spite of the strong chaotic behavior, tidal effects acting in the \stc system are able to stabilise it.
However, the final semi-major axis of the two orbits are $a_1 = 528$~km and $a_2 = 11\,966$~km, which means that they are respectively smaller and higher than the initial and presently observed values $a_1 = 807$~km and $a_2 = 7\,445$~km.
The final outcome of the simulations cannot therefore be interpreted as a possible representation of the present system.
We could try to set higher and/or lower values for the initial semi-major axes, and see if any of the modified tentative systems ended with semi-major axes similar to today's observations.
This exercise is nevertheless not straightforward, because each initial condition has unpredictable behavior while in the chaotic zone.

In section~\ref{initorb} we saw that the present best fitted orbits may not be completely correct.
Stable solutions for the \stc system are possible if we chose a smaller value of the inner orbit eccentricity and a longitude of the pericenter close to $130^\circ$.
In a second experiment, we run a simulation with and without tidal effects for the present best guess for the \stc system parameters (Table\,\ref{tabBen}), but adopting an initial inner orbit eccentricity $e_1 = 0.03$ and longitude of the pericenter $\varpi_1 = 130^\circ$.
The initial rotation period of the two inner bodies is synchronous, the rotation of the outer body is 12~h, and the initial obliquities are $5^\circ$.
In Figure~\,\ref{tides2} we show the evolution of the semi-major axes and eccentricities.
We now observe that tides almost do not change the system, 
since the modified initial conditions already correspond to a system that has been damped by tides.
The modified initial solution is thus a more realistic representation of the true system.

\begin{figure*}
\begin{center}
\includegraphics[width=\textwidth]{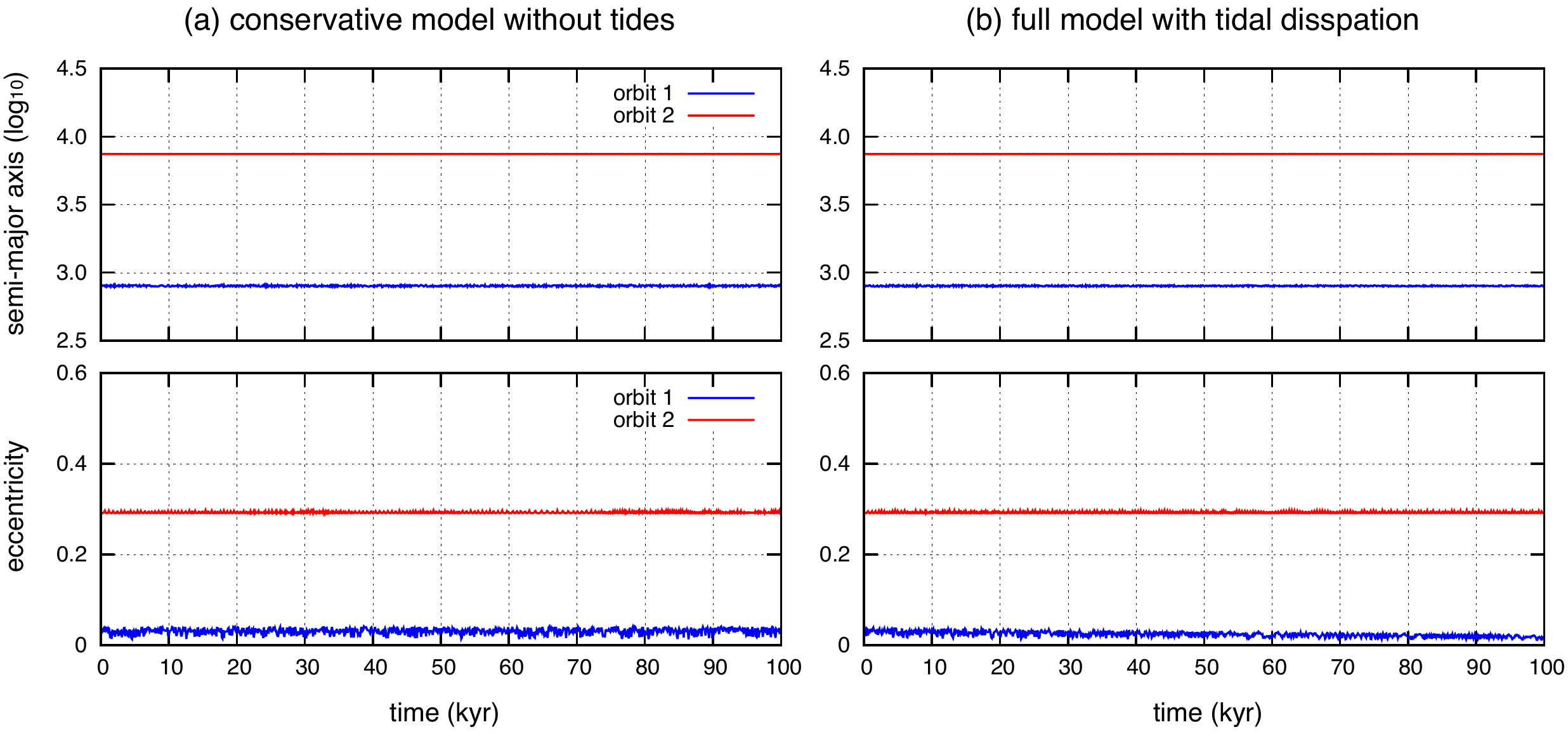} 
 \caption{Orbital evolution of the \stc system (Table\,\ref{tabBen}) with initial $e_1=0.03$ and $\varpi_1 = 130^\circ$ over $10^5$~yr with (a) and without (b) tidal effects.
The initial rotation period of the two inner bodies is synchronous, the rotation of the outer body is 12~h, and the initial obliquities are $5^\circ$. In the top panel we show the semi-major axes $\log_{10} a_{[\mathrm{km}]}$, while in the bottom panel we show the eccentricities. In this case tidal dissipation has almost no impact in the evolution.
 \llabel{tides2}  }
\end{center}
\end{figure*}

\section{Discussion}

The dynamics of the \stc system presents a large number of possible behaviors.
In particular, the system can show chaotic variations for the spins and orbits for a wide range of its parameters.
A more precise analysis of this system requires a better knowledge of the present orbits, spin states and shapes of all bodies involved.

A major conclusion of our analysis is that for realistic values of the shapes and spin states, the present orbits are unstable, mainly because the eccentricity of the inner orbit is too high.
It is then possible that the present eccentricity is incorrectly determined.
As observed in \citet{Benecchi_etal_2010}, due to a short circularisation time-scale ``the inner pair would be expected to have evolved to zero eccentricity''.
Indeed, we have included tidal effects in our simulations and see that the eccentricity of the inner orbit is reduced to very small values in less than 100~kyr.

The best-fitted orbital solution listed in Table\,\ref{tabBen} was obtained using simple point-mass non-interacting Keplerian orbits \citep{Benecchi_etal_2010}.
However, mutual perturbations and spin-orbit interactions excite the eccentricities of both orbits on very short time-scales, shorter than the time span of the already existing observations (7~yr).
Indeed, \citet{Benecchi_etal_2010} already reported that the reduced $\chi^2$ increased when they add data points farther away in time, which is expected when the system is not Keplerian.

A better fit to the observational data would require a more complete model, like the one described in section~\ref{nbody}.
Over the course of the present day observations the conservative model is enough, because tides only become efficient after thousands of years.
Nevertheless, we can improve our adjustment to take indirectly into account the effects of tides, by forcing the amplitude of the inner orbit main proper mode of the eccentricity to be zero \citep[for details see][]{Laskar_etal_2012}.
By doing this kind of indirect adjustment we are able to find the correct combination of ($e_1, \varpi_1$) that stabilises the system for a given ($e_2, \varpi_2$), as in the example shown in Figure~\ref{figew}.

For simplicity, in all simulations shown in this work we adopted an initial obliquity $\theta = 5^\circ$ for all bodies.
This choice was motivated in one hand because tidal effects also damp the obliquities to very small values \citep[e.g.][]{Hut_1980, Correia_Laskar_2010B}, and on the other hand because for trajectories in the chaotic zone the initial obliquity is not very important (see Figures~\ref{fig21} and~\ref{fig22}).
We have nevertheless run some simulations with larger values for the initial obliquity.
As in previous studies \citep[e.g.][]{Correia_etal_2015}, we observed that the size of the chaotic zones increased.
The results in this paper on the extent of the chaotic areas can therefore be seen as conservative, i.e., for larger initial obliquities we can expect larger chaotic areas.

\bfx{ 
In all simulations we adopted the presently observed inclinations and longitude of nodes (Table\,\ref{tabBen}). 
This solution corresponds to a system with an initial mutual inclination of $10.7^\circ$. 
We never show the inclination evolution in our simulations because its behaviour is not very interesting.
For stable configurations the inclination only undergoes small oscillations around the initial value.
For chaotic orbits the mutual inclination is also chaotic and can present erratic motion within $0^\circ$ and about $20^\circ$.}

\bfx{
In our model, we neglected terms in $(R/r)^3$ in the expression of the gravitational potencial (\ref{121026b}), i.e., we adopted a quadrupolar approximation for the spin-orbit interactions.
For irregularly shaped bodies orbiting so close to each other, higher order terms may become important \citep[e.g.][]{Batygin_Morbidelli_2015, Boue_2017}.
However, given the present incertitudes in the shapes, rotational periods and orbits for the \stc system, the quadrupole approximation already gives a very global picture on its dynamics.
We expect that additional corrections in the gravitational potential may extend the sizes of the chaotic areas.
}

The \stc system has been proposed as a possible target for the NASA {\it New Horizons 2} mission \citep[see Ch. 8 in][]{Czysz_Bruno_2006}, which was not selected for further development.
However, we have seen that the \stc system is among the most interesting  trans-Neptunian objects, not only for the present dynamical stability, but also because its formation is very challenging.
Therefore, we encourage the scientific community to include again the \stc system as a possible target for future missions to the Kuiper-belt.

\section*{Acknowledgments}

We thank A. Leleu, N. Peixinho and P. Robutel for discussions.
We acknowledge support from CIDMA strategic project UID/MAT/04106/2013, 
and from 
ENGAGE SKA, POCI-01-0145-FEDER-022217, funded by COMPETE 2020 and FCT, Portugal.


\bibliographystyle{elsarticle-harv}      
\bibliography{correia}

\end{document}